\RequirePackage{fix-cm}
\documentclass[twocolumn]{svjour3}          
\smartqed  
\usepackage{graphicx}
\usepackage[utf8]{inputenc} 
\usepackage{graphicx}
\usepackage{color}
\usepackage{amsmath}
\usepackage{amsfonts}
\usepackage{float}
\usepackage{pstricks}
\usepackage{subfig}
\usepackage{wrapfig}
\usepackage[a4paper]{geometry}
\usepackage{enumerate}
\usepackage{natbib}

\newcommand{\bfX}{{\bf X}}

\newcommand{\bfU}{{\bf U}}
\newcommand{\bfu}{{\bf u}}
\newcommand{\bfv}{{\bf v}}
\newcommand{\bfz}{{\bf z}}
\newcommand{\bfx}{{\bf x}}
\newcommand{\bfZ}{{\bf Z}}

\newcommand{\bfB}{{\bf B}}

\newcommand{\bfR}{{\bf R}}
\newcommand{\bfr}{{\bf r}}
\newcommand{\bfq}{{\bf q}}
\newcommand{\bfV}{{\bf V}}

\begin{document}

\title{Regularised PCA to denoise and visualise data
}


\author{Marie Verbanck         \and
        Julie Josse \and
        François Husson 
        }


\institute{M. Verbanck \at
              Applied mathematics department, Agrocampus Ouest\\
              Tel.: +332-23-48-54-91\\
              \email{marie.verbanck@agrocampus-ouest.fr}           
           \and
           J. Josse \at
              Applied mathematics department, Agrocampus Ouest\\
              Tel.: +332-23-48-58-74\\
              \email{julie.josse@agrocampus-ouest.fr}           
           \and
           F. Husson \at
              Applied mathematics department, Agrocampus Ouest\\
              Tel.: +332-23-48-58-86\\
              \email{francois.husson@agrocampus-ouest.fr}           
}

\date{Received: date / Accepted: date}

\maketitle

\begin{abstract}
Principal component analysis \linebreak (PCA) is a well-established dimensionality reduction method commonly used to denoise and visualise data. A classical PCA model is the fixed effect model in which data are generated as a fixed structure of low rank corrupted by noise. Under this model, PCA does not provide the best recovery of the underlying signal in terms of mean squared error. 
Following the same principle as in ridge regression, we suggest a regularised version of PCA that 
essentially selects a certain number of dimensions and shrinks the corresponding singular values.
Each  singular value is multiplied by a term which can be seen as the ratio of the signal variance over the total variance of the associated dimension.
The regularised term is analytically derived using asymptotic results and can also be justified from a Bayesian treatment of the model.
Regularised PCA provides promising results in terms of the recovery of the true signal and the graphical outputs in comparison with classical PCA and with a soft thresholding estimation strategy.
The distinction between PCA and regularised PCA becomes especially important in the case of very noisy data.

\end{abstract}


\section{Introduction}
\label{introduction}

In many applications \citep{mazumder2010, candes2012}, we can consider that data are generated as a structure having a low rank representation corrupted by noise. Thus, the associated model for any data matrix $\bfX$ (assumed without loss of generality to be centered) composed of $n$ individuals and $p$ variables can be written as:
\begin{eqnarray}
\bfX_{n \times p} & = & \tilde\bfX_{n \times p} + \varepsilon_{n \times p} \label{EQmodel1}\\
x_{ij}  &=& \sum_{s = 1}^{S} \sqrt{d_s} q_{is} r_{js}   + \varepsilon_{ij} 
\mbox {, }  \ \varepsilon_{ij} \sim \mathcal{N}(0, \sigma^2) \nonumber
\end{eqnarray}
where $d_s$ is the $s^{th}$ eigenvalue of the matrix $\tilde{\bfX}'\tilde{\bfX}$ ($n$ times the true covariance matrix), $\bfr_s = \{r_{1s}, ..., r_{js}, ..., r_{ps}\}$ is the associated eigenvector and $\bfq_s = \{q_{1s}, ..., q_{is}, ..., q_{ns}\}$ is the $s^{th}$ eigenvector of the matrix $\tilde{\bfX}\tilde{\bfX}'$ ($n$ times the true inner-product matrix).
Such a model is also known as the fixed effect model \citep{caussinus1986} in principal component analysis \linebreak (PCA).

PCA is a well-established dimensionality reduction method. 
It allows the data $\bfX$ to be described using a small number ($S$) of uncorrelated variables (the principal components) while retaining as much information as possible. PCA is often used as an exploratory \linebreak method to summarise and visualise data.
PCA is also often considered as a way of separating the signal from the noise where the first $S$ principal components are taken as the signal while the remaining ones  as the noise. 
Therefore, PCA can be used as a denoising method to analyse images for instance or to preprocess data before applying other methods such as clustering. Indeed, clustering is expected to be more stable when applied to noise-free data sets.

PCA provides a subspace which best represents the data, that is, which minimises the distances between individuals and their projection on the subspace. Formally, this corresponds to finding a matrix $\hat \bfX_{n \times p}$, of low rank $S$, which minimises $\|\bfX - \hat \bfX \|^2$ with $\|\bullet\|$ the Frobenius norm.
The solution is given by the singular value decomposition (SVD) of $\bfX$: 
\begin{eqnarray}
\hat x_{ij}  = \sum_{s = 1}^{S} \sqrt{\lambda_s} u_{is} v_{js}
\label{mv}
\end{eqnarray}
where $\lambda_s$ is the $s^{th}$ eigenvalue of $\bfX'\bfX$, $\bfu_s = \{u_{1s}, ..., u_{is}, ..., u_{ns}\}$  the $s^{th}$ left singular vector and $\bfv_s = \{v_{1s}, ..., v_{js}, ..., v_{ps}\}$  the $s^{th}$ right singular vector.
This least squares estimator corresponds to the maximum likelihood solution of model \eqref{EQmodel1}.

It is established, for instance in regression, that the maximum likelihood estimators are not necessarily the best for minimising mean squared error (MSE).
However, shrinkage estimators, although biased, have smaller variance which may reduce the MSE. We follow this approach and propose a regularised version of PCA in order to get a better estimate of the underlying structure $\tilde \bfX$. 
In addition, this approach allows graphical representations which are as close as possible to the representations that would be obtained from the signal only.
As we will show later, our approach essentially shrinks the first $S$ singular values with a different amount of shrinkage for each singular value. The shrinkage terms will be analytically derived.

In the literature, a popular strategy to recover a low rank signal from noisy data is to use a soft thresholding strategy. More precisely, each singular value is thresholded with a constant amount of shrinkage usually found by cross-validation.
However, recently, \citet{candes2012} suggested determining the threshold level without resorting to a computational method by minimising an estimate of the risk, namely a Stein's unbiased risk estimate (SURE). 
We will compare our approach to this SURE method.

In this paper, we derive the shrinkage terms by minimising the mean squared error and define regularised PCA (rPCA) in Section \ref{sec:rPCA}.
We also show that rPCA can be derived from a Bayesian treatment of the fixed effect model \eqref{EQmodel1}.
Section \ref{sec:performance} shows the efficiency of regularisation through a simulation study in which rPCA is compared to classical PCA and the \linebreak SURE method. The performance of rPCA is illustrated through the recovery of the signal and the graphical outputs (individual and variable representations). Finally, rPCA is \linebreak performed on a real microarray data set and on images in Section \ref{sec:application}.

\section{Regularised PCA}
\label{sec:rPCA}
				\subsection{MSE point of view}
				\label{sec:MSE}
				
				\subsubsection{Minimising the MSE}		
				\label{sec:minMSE}

PCA provides  an estimator $\hat{\bfX}$ which is as close as possible to $\bfX$ in the least squares sense. 
However, assuming model \eqref{EQmodel1}, the objective is to get an estimator as close as possible to the unknown signal $\tilde{\bfX}$.
To achieve such a goal, the same principle as in ridge regression is followed. We look for a shrinkage version of the maximum likelihood estimator which is as close as possible to the true structure. More precisely, we look for shrinkage terms \linebreak  ${\bf \Phi} = (\phi_s)_{s=1,...,\min(n-1, p)}$ that minimise:
\begin{eqnarray*}
&&\mbox{MSE}  =  \mathbb{E} \left(\sum_{i, j}  \left(\sum_{s = 1}^{\min(n-1, p)} \phi_s \hat{x}_{ij}^{(s)} - \tilde{x}_{ij}^{(s)} \right)^2\right) \\  &&\text{ with } \hat{x}_{ij}^{(s)} = \sqrt{\lambda_s} u_{is} v_{js} \mbox{; } \tilde{x}_{ij}^{(s)}=\sqrt{d_s} q_{is} r_{js} 
\end{eqnarray*}
First, we separate the terms of the MSE corresponding to the first $S$ dimensions from the remaining ones:
\begin{eqnarray*}
\mbox{MSE} & = & \mathbb{E} \left(\sum_{i, j} 
\left(\sum_{s = 1}^{S} \phi_s \hat{x}_{ij}^{(s)} - \tilde{x}_{ij}^{(s)} \right)^2 +  \right.\\
& &  \left.\left(\sum_{s = S+1}^{\min(n-1, p)} \phi_s \hat{x}_{ij}^{(s)} - \tilde{x}_{ij}^{(s)} \right)^2
\right)
\end{eqnarray*}
Then, according to equation \eqref{EQmodel1}, for all $s \geq S+1$, $\tilde{x}_{ij}^{(s)} = 0$. Therefore, the MSE is minimised for $\phi_{S+1} = ... = \phi_{\min(n-1, p)} = 0$.
Thus, the MSE can be written as:
\begin{eqnarray*}
\mbox{MSE} & = & \mathbb{E} \left(\sum_{i, j} 
\left(\sum_{s = 1}^{S} \phi_s \hat{x}_{ij}^{(s)} - \tilde{x}_{ij}^{(s)} \right)^2
\right)
\end{eqnarray*}
Using the orthogonality constraints, for all $s \neq s', \sum_i u_{is} u_{is'} = \sum_j v_{js} v_{js'} =0$, the MSE can be simplified as follows:
\begin{eqnarray}
\mbox{MSE} &= &  \mathbb{E} \left(\sum_{i, j} \left(\sum_{s = 1}^S \phi_s^2 \lambda_s u_{is}^2 v_{js}^2 \right.\right.  \nonumber\\
& - &\left.\left. 2 \tilde{x}_{ij}  \sum_{s = 1}^S \phi_s \sqrt{\lambda_s} u_{is} v_{js} + (\tilde{x}_{ij})^{2} \right)\right)
\label{EQ_mse}
\end{eqnarray}
Finally, equation~\eqref{EQ_mse} is differentiated with respect to $\phi_s$ to get:
\begin{eqnarray*}	
\phi_s  &= &\frac{\sum_{i, j} \mathbb{E}\left(\hat{x}^{(s)}_{ij}\right) \tilde{x}_{ij}}{\sum_{i, j} \mathbb{E}\left(\hat{x}_{ij}^{(s)2}\right)}\\
& =& \frac{\sum_{i, j} \mathbb{E}\left(\hat{x}^{(s)}_{ij}\right) \tilde{x}_{ij}}{\sum_{i, j} \left(\mathbb{V}\left(\hat{x}_{ij}^{(s)}\right) + \left(\mathbb{E}\left(\hat{x}_{ij}^{(s)}\right)\right)^2 \right)}
\end{eqnarray*}

Then, to simplify this quantity, we adapt results coming from the setup of analysis of variance with two factors to the PCA framework. More precisely, we use the results of  \cite{pazman1999} and \cite{denis1996} who studied nonlinear regression models with constraints and focused on bilinear models called biadditive models. Such models are defined as follow:
\begin{eqnarray}
y_{ij}=\mu+\alpha_i+\beta_j+\sum_{s=1}^S \gamma_{is}\delta_{js} + \varepsilon_{ij} \label{biadditive}\\
~~~~~ \mbox{with}~~\varepsilon_{ij}\sim {\cal N}(0,\sigma^2) \nonumber
\end{eqnarray}
where 
$y_{ij}$ is the response for the category $i$ of the first factor and the category $j$ of the second factor,
$\mu$ is the grand mean, $(\alpha_i)_{i=1,...,I}$ and $(\beta_j)_{j=1,...,J}$ correspond to the main effect parameters and $\left(\sum_{s=1}^S\gamma_{is}\delta_{js}\right)_{i=1,...,I ; j=1,...,J}$ model the interaction. The least squares estimates of the multiplicative terms are given by the singular value decomposition of the residual matrix of the model without interaction. From a computational point of view,
this model is similar to the PCA one, the main difference being that the
linear part only includes the grand mean and column main effect in PCA.
Using the Jacobians and the Hessians of the response defined by \citet{denis1994} and recently in \citet{papadopoulo2000}, \citet{denis1996} derived the asymptotic bias of the response of model \eqref{biadditive} and showed that the response estimator is approximately unbiased. Transposed to the PCA framework, it leads to conclude that the PCA estimator is asymptotically unbiased $\mathbb{E}\left(\hat{x}_{ij}\right) = \tilde{x}_{ij}$ and for each dimension $s$, $\mathbb{E}\left(\hat{x}^{(s)}_{ij}\right) = \tilde{x}^{(s)}_{ij}$.
In addition, the variance of $\hat{x}_{ij}$ can be approximated by the noise variance.
Therefore, we estimate $\mathbb{V}\left(\hat{x}_{ij}^{(s)}\right)$ by the average variance, that is $\mathbb{V}\left(\hat{x}_{ij}^{(s)}\right) = \frac{1}{\min(n-1; p)}  \sigma^2$.\\


Consequently $\phi_s$ can be approximated by: 
\begin{equation*}
\phi_s  = \frac{\sum_{i, j} \tilde{x}^{(s)}_{ij} \tilde{x}_{ij}}{\sum_{i, j} \left( \frac{1}{\min(n-1; p)}\sigma^2 + (\tilde{x}^{(s)}_{ij})^2 \right)}
\end{equation*}
Since for all $s \ne s'$, the dimensions $s$ and $s'$ of $\tilde{\bfX}$ are orthogonal, thus $\phi_s$ can be written as:
\begin{equation*}
\phi_s  = 	\frac{\sum_{i, j} \tilde{x}_{ij}^{(s)2}}{\sum_{i, j} \left( \frac{1}{\min(n-1; p)}\sigma^2 + (\tilde{x}^{(s)}_{ij})^2 \right)}
\end{equation*}     
Based on equation \eqref{EQmodel1}, the quantity $\sum_{i, j} (\tilde{x}_{ij}^{(s)})^2$ is equal to $d_s$ the variance of the $s^{th}$ dimension of the signal. $\phi_s$ is then equal to:
\begin{equation}
\phi_s  
= 
  \left\{
      \begin{aligned}
      \frac{d_s}{\frac{np}{min\{p, n-1\}} \sigma^2 + d_s} \forall s = 1, ..., S\\
	0 ~~~~~~~~~~~~~~~~~~~~~~~~~~~\mbox{otherwise}
      \end{aligned}
    \right.
\label{shrinkage_factor}
\end{equation}   
The form of the shrinkage term is appealing since it corresponds to the ratio of the variance of the signal over the total variance (signal plus noise) for the $s^{th}$ dimension.\\


\noindent \textit{Remark: Models such as model \eqref{biadditive} are also known as additive main effects and multiplicative interaction (AMMI) models. They are often used to analyse genotype-environment data in plant breading framework. Considering a random version of such models, \linebreak \citet{cornelius1999} developed a regularisation term which is similar to ours. It allows improved prediction of the yield obtained by genotypes in environments}


\subsubsection{Definition of regularised PCA}
\label{sec:rPCA_exp}

The shrinkage terms \eqref{shrinkage_factor} depend on unknown quantities. We estimate them by plug-in.
The total variance of the $s^{th}$ dimension is estimated by the variance of $\bf X$ for the dimension $s$, \emph{i.e.} by its associated eigenvalue $\lambda_s$.
The signal variance of the $s^{th}$ dimension is estimated by the estimated total variance of the $s^{th}$ dimension minus an estimate of the noise variance of the $s^{th}$ dimension.
Consequently, $\phi_s$ is estimated by $\hat \phi_s=\frac{\lambda_s-  \frac{np}{\min(n-1; p)}\hat \sigma^2}{\lambda_s}$.
Regularised PCA \linebreak (rPCA) is thus defined by multiplying the maximum likelihood solution by the shrinkage terms which leads to:
\begin{eqnarray}
\hat{x}_{ij}^{\mbox{rPCA}}&&= \sum_{s=1}^{S} \left(\frac{\lambda_s-\frac{np}{\min(n-1; p)}\hat\sigma^2}{\lambda_s}\right)\sqrt{\lambda_s} u_{is} v_{js} \nonumber\\
= &&\sum_{s=1}^{S} \left(\sqrt{\lambda_s}-\frac{\frac{np}{\min(n-1; p)}\hat \sigma^2}{\sqrt{\lambda_s}}\right) u_{is} v_{js} \label{rPCAcellule}
\end{eqnarray}
Using matrix notations, with $\bfU$ being the matrix of the first $S$ left singular vectors of $\bf X$, $\bfV$ being the matrix of the first $S$ right singular vectors of $\bf X$ and $\bf \Lambda$ being the diagonal matrix with the associated eigenvalues, the fitted matrix by rPCA is:
\begin{eqnarray}
\hat{\bfX}^{\mbox{rPCA}}&& = \bfU {\bf \hat{\Phi}} {\bf \Lambda}^{1/2} \bfV' \label{rPCAmatrix}
\end{eqnarray}
rPCA essentially shrinks the first $S$ singular values. It can be interpreted as a compromise between hard and soft thresholding. 
Hard thresholding consists in selecting a certain number of dimensions $S$ which corresponds to classical PCA (equation \ref{mv}) whereas soft \linebreak thresholding consists in thresholding all singular values with the same amount of shrinkage (and without prespecifying the number of dimensions).
In rPCA, the $s^{th}$ singular value is less shrunk than the $(s+1)^{th}$ one.
This can be interpreted as granting a greater weight to the first dimensions. 
This behaviour seems desirable. 
Indeed, the first dimensions can be considered as more stable and trustworthy than the last ones. 
The regularisation procedure relies more heavily on the less variable dimensions.
When $\hat{\sigma}^2$ is small, $\hat{\phi}_s$ is close to 1 and rPCA reduces to standard PCA. 
When $\hat{\sigma}^2$ is high, $\hat{\phi}_s$ is close to 0 and the values of $\hat{\bfX}^{\mbox{rPCA}}$ are close to 0 which corresponds to the average of the variables (in the centered case).
From a geometrical point of view, rPCA leads to bring the individuals closer to the centre of gravity.

The regularisation procedure requires estimation of the residual variance $\sigma^2$. As the maximum likelihood estimator is biased, another estimator corresponds to the ratio of the residual sum of squares divided by the number of observations minus the number of independent parameters. The latter are equal to \linebreak $p + \left((nS - S) - \frac{S(S+1)}{2}\right) +  \left(pS - \frac{S(S+1)}{2} - S\right)$,
\emph{i.e.} $p$ parameters for the centering, \linebreak $\left((nS - S) - \frac{S(S+1)}{2}\right)$ for the centered and orthonormal left singular vectors and \linebreak $\left(pS - \frac{S(S+1)}{2} - S\right)$ for the orthonormal right singular vectors. 
This number of parameters can also be calculated as the trace of the projection matrix involved in PCA \citep{Candes09,josse2011}.
Therefore, the residual variance is estimated as:
\begin{eqnarray}
\hat{\sigma}^2 & = & \frac{\| \bfX - \hat \bfX \|^2}{np - p - nS - pS + S^2 +S} \nonumber\\
 & = & \frac{\sum_{s = S+1}^{\min(n-1; p)} \lambda_s}{np - p - nS - pS + S^2 +S}
\end{eqnarray}
Contrary to many methods, this classical estimator, namely the residual sum of squares divided by the number of observations minus the number of independent parameters, is still biased. 
This can be explained by the non-linear form of the model or by the fact that the projection matrix \citep{josse2011} depends on the data.


\subsection{Bayesian points of view}
\label{sec:Bayesian}


Regularised PCA has been presented and defined via the minimisation of the MSE in section \ref{sec:MSE}. However, it is possible to define the method without any reference to MSE, instead using Bayesian considerations. It is well known, in linear regression for instance, that there is equivalence between ridge regression and a \linebreak Bayesian treatment of the regression model. More precisely, the maximum a posteriori of the regression parameters assuming a Gaussian prior for these parameters corresponds to the ridge estimators \citep[p. 64]{hastie2009}.  
Following the same rationale, we suggest in this section a Bayesian treatment of the fixed effect model \eqref{EQmodel1}.

First, several comments can be made on this model. It is called a ``fixed effect" model since the structure is considered fixed.
Individuals have different expectations and randomness is only due to the error term. This model is most justified in situations where PCA is performed on data in which the individuals themselves are of interest and are not a random sample drawn from a population of individuals. Such situations frequently arise in practice. 
For instance, in sensory analysis, individuals can be products, such as chocolates, and variables can be sensory descriptors, such as bitterness, sweetness, etc. The aim is to study these specific products and not others (they are not interchangeable). It thus makes sense to estimate the individual parameters ($\bfq_s$) and to study the graphical representation of the individuals as well as the representation of the variables.
In addition, let us point out that the inferential framework associated with this model is not usual. Indeed the number of parameters increases when the number of individuals increases. Consequently, in this model, asymptotic results are obtained by considering that the noise variance tend to 0.


To suggest a Bayesian treatment of the fixed effect model, we first recall the principle of probabilistic PCA \citep{roweis1998,tipping1999} which will be interpreted as a Bayesian treatment of this model.

\subsubsection{Probabilistic PCA model}
\label{sec:ACPP}

The probabilistic PCA (pPCA) model is a particular case of a factor analysis model \linebreak \citep{bartholomew1987} with an isotropic noise. The idea behind these models is to summarise the relationships between variables using a small number of latent variables.  More precisely, denoting $\bfx_i$ a row of the matrix $\bfX$, the pPCA model is written as follows:
\begin{eqnarray*}
\bfx_i & = & \bfB_{p \times S} \bfz_i + \varepsilon_i  \\ && \bfz_i \sim \mathcal{N}(0, \mathbb{I}_S) \text{, } \varepsilon_i \sim \mathcal{N}(0, \sigma^2 \mathbb{I}_p)
\label{mod_acpp}
\end{eqnarray*}
with $\bfB_{p \times S}$ being the matrix of unknown coefficients, $\bfz_i$ being the latent variables and $\mathbb{I}_S$ and $\mathbb{I}_p$ being the identity matrices of size $S$ and $p$.
This model induces a Gaussian distribution on the individuals (which are independent and identically distributed) with a specific structure of variance-covariance:
\begin{eqnarray*}
\bfx_i  \sim \mathcal{N}(0, {\bf \Sigma}) \text{ with } {\bf \Sigma} = \bfB\bfB' + \sigma^2 \mathbb{I}_p
\end{eqnarray*}

There is an explicit solution for the maximum likelihood estimators:
\begin{eqnarray}
\hat{\bfB}= \bfV ({\bf\Lambda} - \sigma^2 \mathbb{I}_S)^\frac{1}{2} \bfR \mbox{  } \label{EQw}
\end{eqnarray}
with $\bfV$ and $\bf \Lambda$ defined as in equation \eqref{rPCAmatrix}, that is, as the matrix of the first $S$ left singular vectors of $\bf X$ and as the diagonal matrix of the eigenvalues, $\bfR_{S \times S}$ a rotation matrix (usually equal to $\mathbb{I}_S$) and $\sigma^2$ estimated as the mean of the last eigenvalues.

In contrast to the fixed effect model \eqref{EQmodel1}, the pPCA model can be seen as a random effect model since the structure is random because of the Gaussian distribution on the latent variables. Consequently, this model seems more appropriate when PCA is performed on sample data such as survey data. 
In such cases,  the individuals are not themselves of interest but only considered for the information they provide on the links between variables.  Consequently, in such studies, at first, it does not make sense to consider ``estimates'' of the ``individual parameters'' since no parameter is associated with the individuals, only random variables ($\bfz_i$). 
However, estimators of the ``individual parameters'' are usually calculated as the expectation of the latent variables given the observed variables $\mathbb{E}(\bfz_i|\bfx_i)$. The calculation is detailed in \citet{tipping1999} and results in:
\begin{eqnarray}
\hat \bfZ= \bfX \hat{\bfB}(\hat{\bfB}'\hat{\bfB}+\sigma^2\mathbb{I}_S)^{-1}
\label{esp_cond}
\end{eqnarray}
We can note that such estimators are often called BLUP estimators \citep{robinson1991} in the framework of mixed effect models where it is also customary to give estimates of the random effects.
 
Thus, using the maximum likelihood estimator of $\bfB$ (equation \ref{EQw}) and equation \eqref{esp_cond}, it is possible to build a fitted matrix as:
\begin{eqnarray*}
\hat{\bfX}^{\mbox{pPCA}} & = & \hat{\bfZ} \hat{\bfB}'  = \bfX \hat{\bfB}(\hat{\bfB}'\hat{\bfB}+\sigma^2\mathbb{I}_S)^{-1}\hat{\bfB}' \\
& = & \bfX \bfV ({\bf\Lambda} - \sigma^2\mathbb{I}_S)^\frac{1}{2} {\bf\Lambda}^{-1} ({\bf\Lambda} - \sigma^2\mathbb{I}_S)^\frac{1}{2} \bfV' \\
&=& \bfU ({\bf\Lambda} - \sigma^2\mathbb{I}_S) {\bf\Lambda}^{-\frac{1}{2}} \bfV' \\ \mbox{~since~} \bfX &\bfV &={\bf\Lambda}^{1/2}\bfU \mbox{~(given by the SVD of \bfX)~}
\end{eqnarray*}
Therefore, considering the pPCA model  leads to a fitted matrix of the same form as $\hat\bfX^{\mbox{rPCA}}$ defined in equation \eqref{rPCAmatrix} with the same shrunk singular values $\left({\bf\Lambda} - \sigma^2\mathbb{I}_S\right) {\bf\Lambda}^{-1/2}$.
However, the main difference between the two approaches is that the pPCA model considers individuals as random, whereas they are fixed in model~\eqref{EQmodel1} used to define rPCA. 
Nevertheless, from a conceptual point of view, the random effect model can be considered as a Bayesian treatment of the fixed effect model with a prior distribution on the left singular vectors.
Thus, we can consider the pPCA model as the fixed effect model on which we assume a distribution on $\bfz_i$, considered as the ``individual parameters''. It is a way to define constraints on the individuals. 

\noindent \emph{Remark: Even if a maximum likelihood solution is available (equation~\ref{EQw}) in pPCA, it is possible to use an EM algorithm \citep{rubin1982} to estimate the parameters. The two steps correspond to the following two multiple ridge regressions:
\begin{eqnarray*}
\text{Step E: } & & \hat{\bfZ} = \bfX \hat{\bfB} (\hat{\bfB}'\hat{\bfB} + \hat{\sigma}^2 \mathbb{I}_S )^{-1}\\
\text{Step M: } & & \hat{\bfB} = \bfX' \hat{\bfZ} (\hat{\bfZ}'\hat{\bfZ} + \hat{\sigma}^2 {\bf\Lambda}^{-1})^{-1}
\end{eqnarray*}
Thus, the link between pPCA and the regularised version of PCA is also apparent in these equations. 
That is, introducing two ridge terms in the two linear multiple regressions which lead to the usual PCA solution (the EM algorithm associated with model~\eqref{EQmodel1} in PCA is also known as the alternative least squares algorithm):
\begin{eqnarray*}
\text{Step E: } & & {\bfU} = \bfX {\bfV} ({\bfV}'{\bfV} )^{-1}\\
\text{Step M: } & & {\bfV} = \bfX' {\bfU} ({\bfU}'{\bfU} )^{-1}
\end{eqnarray*}
}

\subsubsection{An empirical Bayesian approach}
\label{sec:Bayesian_exp}

Another Bayesian interpretation of the regularized PCA can be given considering directly an empirical Bayesian treatment of the fixed effect model with a prior distribution on each cell of the data matrix per dimension: $\tilde{x}_{ij}^{(s)} \sim \mathcal{N}(0,\tau_s^2)$. From model~\eqref{EQmodel1}, this implies that  $x_{ij}^{(s)} \sim \mathcal{N}(0, \tau_s^2 + \frac{1}{\min(n-1;p)}\sigma^2)$. 
The posterior distribution is obtained by combining the likelihood and the priors:
\begin{eqnarray*}
p\left(\tilde x_{ij}^{(s)}|x_{ij}^{(s)}\right) & = & \frac{p\left(x_{ij}^{(s)}|\tilde x_{ij}^{(s)}\right) p\left(\tilde x_{ij}^{(s)}\right)}{p\left(x_{ij}^{(s)}\right)}\\
&=& \frac{
\frac{1}{\sqrt{2\pi \frac{1}{\min(n-1;p)} \sigma^2}} \exp\left[- \frac{\left(x_{ij}^{(s)}-\tilde x_{ij}^{(s)}\right)^2}{2 \frac{1}{\min(n-1;p)} \sigma^2}\right] \times
\frac{1}{\sqrt{2\pi\tau_s^2}} \exp\left[- \frac{\left(\tilde x_{ij}^{(s)}\right)^2}{2  \tau_s^2}\right]
}{\frac{1}{\sqrt{2\pi(\tau_s^2 +\frac{1}{\min(n-1;p)}  \sigma^2)} } \exp\left[- \frac{\left(x_{ij}^{(s)}\right)^2}{2  (\tau_s^2 + \frac{1}{\min(n-1;p)} \sigma^2)}\right]} \\
&=& \frac{1}{\sqrt{2\pi\frac{ \frac{1}{\min(n-1;p)} \sigma^2\tau_s^2}{\frac{1}{\min(n-1;p)} \sigma^2+\tau_s^2}}} \exp\left[- \frac{\left(\tilde x_{ij}^{(s)} - \frac{\tau_s^2}{\tau_s^2+\frac{1}{\min(n-1;p)} \sigma^2}\ x_{ij}^{(s)} \right)^2}{2 \frac{ \frac{1}{\min(n-1;p)} \sigma^2\tau_s^2}{\tau_s^2+\frac{1}{\min(n-1;p)} \sigma^2}}\right]\\
\end{eqnarray*}
The expectation of the posterior distribution is:
\begin{eqnarray*}
\mathbb{E}\left(\tilde x_{ij}^{(s)}|x_{ij}^{(s)}\right) &=& \Phi_s x_{ij}^{(s)} \\
&&\mbox{ with } \Phi_s = \frac{\tau_s^2}{\tau_s^2+\frac{1}{\min(n-1;p)} \sigma^2}
\end{eqnarray*}
This expectation depends on unknown quantities. They are estimated by maximising the likelihood of $\left(x_{ij}^{(s)}\right)_{i=1,...,n; j=1,...,p}$ as a function of $\tau^2_s$ to obtain:
\begin{eqnarray*}
\hat{\tau_s}^2 &=& \left(\frac{1}{np}\lambda_s - \frac{1}{\min(n-1;p)}\hat\sigma^2 \right)
\end{eqnarray*}
Consequently the shrinkage term is estimated as 
$\hat\Phi_s = \frac{\left(\frac{1}{np}\lambda_s - \frac{1}{\min(n-1;p)}\hat\sigma^2 \right)}{ \frac{1}{np} \lambda_s}  = 
\frac{\lambda_s - \frac{np}{\min(n-1;p)}\hat\sigma^2}{\lambda_s}$ 
and also corresponds to the regularisation term \eqref{rPCAcellule} defined in Section \ref{sec:minMSE}.\\

Thus, regularised PCA can be seen as a Bayesian treatment of the fixed effect model with a prior on each dimension.
The variance of the prior is specific to each dimension $s$ and is estimated as the signal variance of the dimension in question ($\lambda_s - \frac{1}{\min(n-1;p)}\hat\sigma^2$).\\

\noindent \textit{Remark: \citet{hoff2007} also proposed a Bayesian treatment of SVD-related models with a primary goal of estimating the number of underlying dimensions. 
Roughly, his proposition consists in putting prior distributions on $\bfU$, $\bf\Lambda$, 
\linebreak\\\\\\\\\\\\\\\\\\\\\\\\\\\\\\\\\\
and $\bfV$. More precisely, he uses von Mises uniform \citep{hoff2009} prior for orthonormal matrices (on the Steifeld manifold \citet{chikuse2003}) for $\bfU$ and $\bfV$ and normal priors for the singular values, forming a prior distribution for the structure $\tilde \bfX$. Then he builds a Gibbs sampler to get draws from the posterior distributions. 
The posterior expectation of $\tilde \bfX$ can be used as a punctual estimate. It can also be seen as a regularised version of the maximum likelihood estimate.
However, contrary to the previously described approach, there is no closed form expression for the regularisation.
}


\subsection{Bias-variance trade-off}
\label{sec:tradeOff}

The rationale behind rPCA can be illustrated on graphical representations. 
Usually, different types of graphical representations are associated with PCA \citep{greenacre_biplots_2010} depending on whether the left and right singular vectors are represented as normed to 1 or to their associated singular value.
In our practice \citep{husson_exploratory_2010}, we represent the individual coordinates by $\bfU {\bf\Lambda}^{\frac{1}{2}}$ and the variable coordinates by  $\bfV {\bf \Lambda}^{\frac{1}{2}}$. Therefore, the global shape of the individual cloud represents the variance. Similarly, in the variable representation, the cosine of the angle between two variables can be interpreted as the covariance.
Since rPCA ultimately modifies the singular values, it will affect both the representation of the individuals and of the variables.
We focus here on the individuals representation.

\begin{figure*}[!ht]
	\center
	\includegraphics[scale = 0.6]{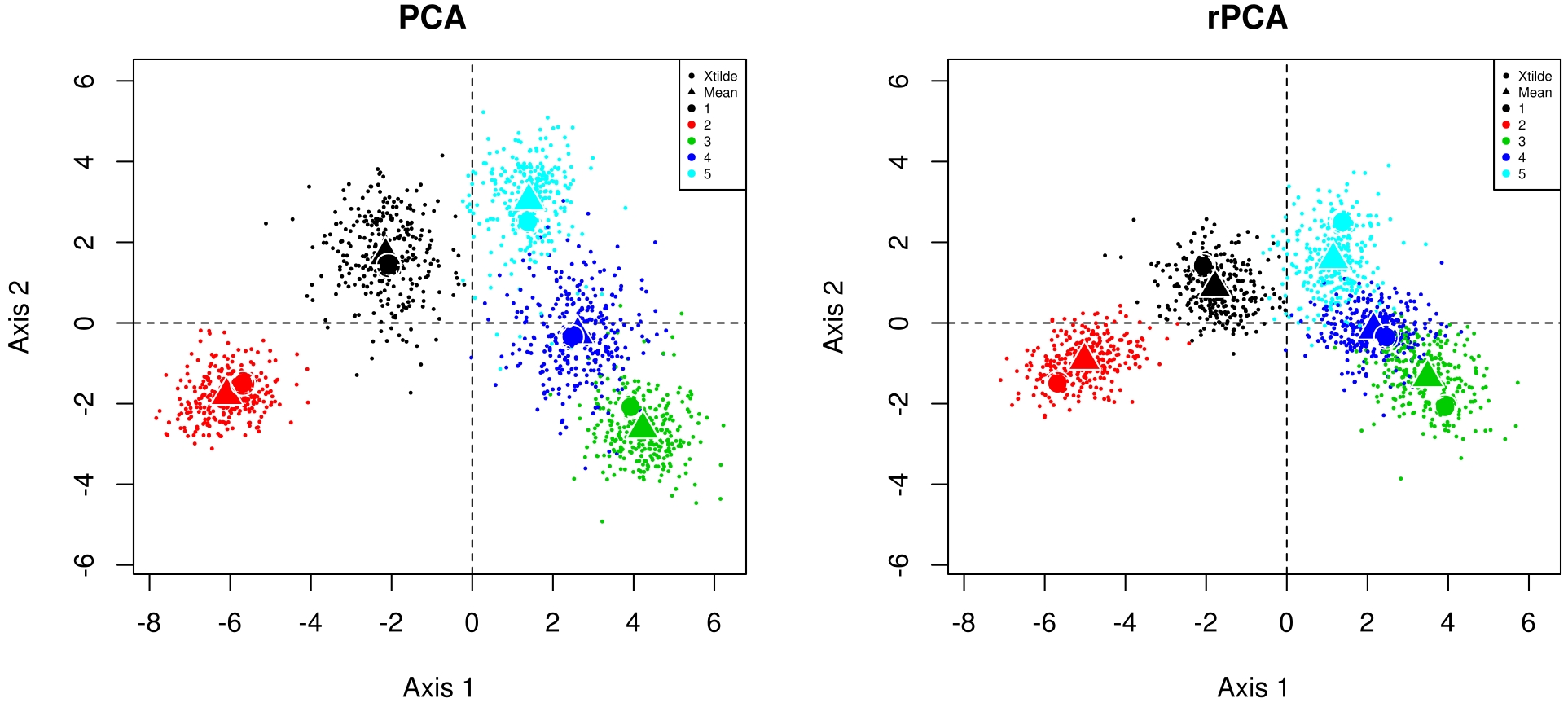}
	\caption{Superimposition of several configurations of individual coordinates using Procustes rotations towards the true individual configuration of $\tilde{\bfX}_{5 \times 15}$ (large dots). Configurations of the PCA (left) and the rPCA (right) of each $\bfX^{sim}= \tilde{\bfX} + \varepsilon^{sim}$, with $sim = 1, ..., 300$ are represented with small dots. The average configuration over the 300 configurations is represented by triangles.}
	\label{ellipse}
\end{figure*}

Data are generated according to model~\eqref{EQmodel1} with an underlying signal $\tilde{\bfX}_{5 \times 15}$ composed of 5 individuals and 15 variables in two dimensions. Then, 300 matrices are generated with the same underlying structure: $\bfX^{sim}=\tilde{\bfX}_{5 \times 15} + \varepsilon^{sim}$ with $sim = 1, ..., 300$. On each data matrix, PCA and rPCA are performed. 
In figure~\ref{ellipse}, the configurations of the 5 individuals obtained after each PCA appear on the left, whereas the configurations obtained after each rPCA appear on the right.
The average configurations over the 300 simulations are represented by triangles and the true individual configuration obtained from $\tilde{\bfX}$ is represented by large dots.
Representing several sets of coordinates from different PCAs can suffer from translation, reflection, dilatation or rotation ambiguities. Thus, all configurations are superimposed using Procustes rotations \citep{gower2004} by taking as the reference the true individuals configuration. 

Compared to PCA, rPCA provides a more biased representation because the coordinates of the average points (triangles) are systematically inferior to the coordinates of the true points (large dots). This is expected because the regularisation term shrinks the individual coordinates towards the origin.
In addition, as it is clear for individual number 4 (dark blue), the representation is less variable. Figure~\ref{ellipse} thus gives a rough idea of the bias-variance trade-off. Note that even the PCA representation is biased, but this is also expected since $\mathbb{E}(\hat{\bfX}) = \tilde{\bfX}$ only asymptotically as detailed in section~\ref{sec:minMSE}.

\section{Simulation study}
\label{sec:performance}

To assess rPCA, a simulation study was conducted and rPCA is compared to classical PCA as well as to the SURE method proposed by \linebreak \citet{candes2012}. As explained in the introduction, the SURE method relies on a soft thresholding strategy: 
\begin{eqnarray*}
\hat{x}_{ij}^{\mbox{SURE}} = \sum_{s=1}^{\min(n,p)} \left(\sqrt{\lambda_s}- \lambda \right)_{+} u_{is} v_{js},
\end{eqnarray*}
The threshold parameter $\lambda$ is automatically selected by minimising Stein's unbiased risk estimate (SURE). 
As a tuning parameter, the SURE method does not require the number of underlying dimensions of the signal, but it does require estimation of the noise variance $\sigma^2$ to determine $\lambda$. 

	\subsection{Recovery of the signal}
	\label{sec:simulationStudy}

Data are simulated according to model~\eqref{EQmodel1}. 
The structure is simulated by varying several parameters:
\begin{itemize}
	\item the number of individuals $n$ and the number of variables	$p$ based on 3 different combinations: ($n=100$ and $p=20$; $n=50$ and $p=50$; $n=20$ and $p=100$)
	\item the number of underlying dimensions $S$ (2; 4)
	\item the ratio of the first eigenvalue to the second eigenvalue $(d_1/d_2)$ of $\tilde{\bfX}$ (4; 1). When the number of underlying dimensions is \linebreak higher than 2, the subsequent eigenvalues are   roughly of the same order of magnitude.
\end{itemize}
More precisely, $\tilde{\bfX}$ is simulated as follows:
\begin{enumerate}
	\item A SVD is performed on a $n \times S$ matrix generated from a standard multivariate normal distribution. The left singular vectors provide $S$ empirically orthonormal vectors.
	\item Each vector $s=1, ..., S$ is replicated to obtain the $p$ variables. The number of times that each vector $s$ is replicated depends on the ratio between the eigenvalues $(d_1/d_2)$. For instance, if $p=50$, $S = 2$, $(d_1/d_2) = 4$, the first vector is replicated 40 times and the second vector is replicated 10 times.
\end{enumerate}
Then, to generate the matrix $\bfX$, a Gaussian isotropic noise is added to the structure. Different levels of variance $\sigma^2$ are considered to obtain three signal-to-noise ratios \citep{mazumder2010} equal to 4, 1 and 0.8. A high signal-to-noise ratio (SNR) implies that the variables of $\bfX$ are very correlated, whereas a low SNR implies that the data are very noisy. For each combination of the parameters, 500 data sets are generated.\\

To assess the recovery of the signal, the MSE is calculated between the fitted matrix $\hat{\bfX}$ obtained from each method and the true underlying signal $\tilde{\bfX}$. The fitted matrices from PCA and rPCA are obtained considering the true number of underlying dimensions as known. The SURE method is performed with the true noise variance as in \citet{candes2012}. Results of the simulation study are gathered in Table~\ref{MSE}.

\begin{table*}[!ht]
\caption{
Mean Squared Error (and its standard deviation) between $\hat{\bfX}$ and $\tilde{\bfX}$ for PCA, rPCA and SURE method over 500 simulations.
Results are given for different numbers of individuals ($n$), numbers of variables ($p$), numbers of underlying dimensions ($S$), signal-to-noise ratios (SNR) and ratios of the first eigenvalue on the second eigenvalue $(d_1/d_2)$.}
\begin{center}
\scriptsize
\begin{tabular}{cccccc|c|c|c}		
& $n$ & $p$ & $S$ & SNR & $(d_1/d_2)$ & $MSE(\hat\bfX^{\mbox{PCA}}, \tilde\bfX)$ & $MSE(\hat\bfX^{\mbox{rPCA}}, \tilde\bfX)$ & $MSE(\hat\bfX^{\mbox{SURE}}, \tilde\bfX)$ \\
\cline{2-9}																														
1	&	100	&	20	&	2	&	4	&	4	&	4.22E-04	(1.69E-06) 	&	\textbf{	4.22E-04	}	(1.69E-06) 	&		8.17E-04		(2.67E-06) 	\\
2	&	100	&	20	&	2	&	4	&	1	&	4.21E-04	(1.75E-06) 	&	\textbf{	4.21E-04	}	(1.75E-06) 	&		8.26E-04		(2.89E-06) 	\\
3	&	100	&	20	&	2	&	1	&	4	&	1.26E-01	(5.29E-04) 	&	\textbf{	1.08E-01	}	(4.56E-04) 	&		1.60E-01		(6.15E-04) 	\\
4	&	100	&	20	&	2	&	1	&	1	&	1.23E-01	(5.05E-04) 	&	\textbf{	1.11E-01	}	(4.61E-04) 	&		1.69E-01		(6.28E-04) 	\\
5	&	100	&	20	&	2	&	0.8	&	4	&	3.34E-01	(1.38E-03) 	&	\textbf{	2.40E-01	}	(9.90E-04) 	&		3.10E-01		(1.05E-03) 	\\
6	&	100	&	20	&	2	&	0.8	&	1	&	3.12E-01	(1.38E-03) 	&	\textbf{	2.45E-01	}	(1.10E-03) 	&		3.32E-01		(1.22E-03) 	\\
7	&	100	&	20	&	4	&	4	&	4	&	8.25E-04	(2.39E-06) 	&	\textbf{	8.24E-04	}	(2.38E-06) 	&		1.42E-03		(3.54E-06) 	\\
8	&	100	&	20	&	4	&	4	&	1	&	8.26E-04	(2.38E-06) 	&	\textbf{	8.25E-04	}	(2.38E-06) 	&		1.43E-03		(3.48E-06) 	\\
9	&	100	&	20	&	4	&	1	&	4	&	2.60E-01	(8.44E-04) 	&	\textbf{	1.96E-01	}	(6.51E-04) 	&		2.43E-01		(6.84E-04) 	\\
10	&	100	&	20	&	4	&	1	&	1	&	2.47E-01	(7.16E-04) 	&	\textbf{	2.04E-01	}	(5.99E-04) 	&		2.62E-01		(6.94E-04) 	\\
11	&	100	&	20	&	4	&	0.8	&	4	&	7.41E-01	(2.69E-03) 	&	\textbf{	4.27E-01	}	(1.53E-03) 	&		4.36E-01		(1.11E-03) 	\\
12	&	100	&	20	&	4	&	0.8	&	1	&	6.68E-01	(2.02E-03) 	&	\textbf{	4.40E-01	}	(1.40E-03) 	&		4.83E-01		(1.33E-03) 	\\
\cline{2-9}																														
13	&	50	&	50	&	2	&	4	&	4	&	2.81E-04	(1.32E-06) 	&	\textbf{	2.81E-04	}	(1.32E-06) 	&		5.95E-04		(2.24E-06) 	\\
14	&	50	&	50	&	2	&	4	&	1	&	2.79E-04	(1.24E-06) 	&	\textbf{	2.79E-04	}	(1.24E-06) 	&		5.93E-04		(2.21E-06) 	\\
15	&	50	&	50	&	2	&	1	&	4	&	8.48E-02	(4.09E-04) 	&	\textbf{	7.82E-02	}	(3.85E-04) 	&		1.26E-01		(4.97E-04) 	\\
16	&	50	&	50	&	2	&	1	&	1	&	8.21E-02	(3.87E-04) 	&	\textbf{	7.77E-02	}	(3.70E-04) 	&		1.31E-01		(5.08E-04) 	\\
17	&	50	&	50	&	2	&	0.8	&	4	&	2.30E-01	(1.12E-03) 	&	\textbf{	1.93E-01	}	(9.64E-04) 	&		2.55E-01		(1.01E-03) 	\\
18	&	50	&	50	&	2	&	0.8	&	1	&	2.14E-01	(9.58E-04) 	&	\textbf{	1.89E-01	}	(8.57E-04) 	&		2.73E-01		(1.07E-03) 	\\
19	&	50	&	50	&	4	&	4	&	4	&	5.48E-04	(1.84E-06) 	&	\textbf{	5.48E-04	}	(1.84E-06) 	&		1.04E-03		(2.82E-06) 	\\
20	&	50	&	50	&	4	&	4	&	1	&	5.46E-04	(1.76E-06) 	&	\textbf{	5.46E-04	}	(1.76E-06) 	&		1.04E-03		(2.79E-06) 	\\
21	&	50	&	50	&	4	&	1	&	4	&	1.75E-01	(6.21E-04) 	&	\textbf{	1.53E-01	}	(5.54E-04) 	&		2.00E-01		(5.79E-04) 	\\
22	&	50	&	50	&	4	&	1	&	1	&	1.68E-01	(5.49E-04) 	&	\textbf{	1.52E-01	}	(5.08E-04) 	&		2.09E-01		(6.04E-04) 	\\
23	&	50	&	50	&	4	&	0.8	&	4	&	5.07E-01	(1.90E-03) 	&		3.87E-01		(1.53E-03) 	&	\textbf{	3.85E-01	}	(1.12E-03) 	\\
24	&	50	&	50	&	4	&	0.8	&	1	&	4.67E-01	(1.62E-03) 	&	\textbf{	3.76E-01	}	(1.38E-03) 	&		4.13E-01		(1.23E-03) 	\\
\cline{2-9}																														
25	&	20	&	100	&	2	&	4	&	4	&	4.22E-04	(1.72E-06) 	&	\textbf{	4.22E-04	}	(1.72E-06) 	&		8.15E-04		(2.80E-06) 	\\
26	&	20	&	100	&	2	&	4	&	1	&	4.21E-04	(1.69E-06) 	&	\textbf{	4.20E-04	}	(1.70E-06) 	&		8.20E-04		(2.89E-06) 	\\
27	&	20	&	100	&	2	&	1	&	4	&	1.25E-01	(5.35E-04) 	&	\textbf{	1.06E-01	}	(4.53E-04) 	&		1.57E-01		(5.83E-04) 	\\
28	&	20	&	100	&	2	&	1	&	1	&	1.22E-01	(5.28E-04) 	&	\textbf{	1.10E-01	}	(4.76E-04) 	&		1.67E-01		(6.20E-04) 	\\
29	&	20	&	100	&	2	&	0.8	&	4	&	3.30E-01	(1.43E-03) 	&	\textbf{	2.35E-01	}	(1.03E-03) 	&		3.06E-01		(1.13E-03) 	\\
30	&	20	&	100	&	2	&	0.8	&	1	&	3.18E-01	(1.30E-03) 	&	\textbf{	2.50E-01	}	(1.03E-03) 	&		3.34E-01		(1.25E-03) 	\\
31	&	20	&	100	&	4	&	4	&	4	&	8.28E-04	(2.38E-06) 	&	\textbf{	8.27E-04	}	(2.39E-06) 	&		1.41E-03		(3.64E-06) 	\\
32	&	20	&	100	&	4	&	4	&	1	&	8.29E-04	(2.58E-06) 	&	\textbf{	8.28E-04	}	(2.58E-06) 	&		1.42E-03		(3.68E-06) 	\\
33	&	20	&	100	&	4	&	1	&	4	&	2.55E-01	(7.59E-04) 	&	\textbf{	1.97E-01	}	(5.92E-04) 	&		2.45E-01		(6.47E-04) 	\\
34	&	20	&	100	&	4	&	1	&	1	&	2.48E-01	(7.45E-04) 	&	\textbf{	2.04E-01	}	(6.20E-04) 	&		2.60E-01		(6.91E-04) 	\\
35	&	20	&	100	&	4	&	0.8	&	4	&	7.13E-01	(2.55E-03) 	&	\textbf{	4.15E-01	}	(1.47E-03) 	&		4.37E-01		(1.19E-03) 	\\
36	&	20	&	100	&	4	&	0.8	&	1	&	6.66E-01	(2.01E-03) 	&	\textbf{	4.34E-01	}	(1.31E-03) 	&		4.78E-01		(1.24E-03) 	\\																											
\end{tabular}
\end{center}
\label{MSE}
\end{table*}

First, rPCA outperforms both PCA and the SURE method in almost all situations. 
As expected, the MSE obtained by PCA and rPCA are roughly of the same order of magnitude when the SNR is high (SNR = 4), as illustrated in rows number 1 or 13, whereas rPCA outperforms PCA when data are noisy (SNR = 0.8) as in rows number 11 or 23. The differences between rPCA and PCA are also more critical when the ratio $(d_1/d_2)$ is high than when the eigenvalues are equal.
When $(d_1/d_2)$ is large, the signal is concentrated on the first dimension whereas it is scattered in more dimensions when the ratio is smaller.
Consequently, the same amount of noise has a greater impact on the second dimension in the first case. This may increase the advantage of rPCA which tends to reduce the impact of noise.

The main characteristic of the SURE \linebreak method observed in all simulations is that it gives particularly good results when the data are very noisy.
Consequently, the results are satisfactory when SNR = 0.8, particularly when the number of underlying dimensions is high (rows number 11, 23 and 35 for instance).  This behaviour can be explained by the fact that the same amount of signal is more impacted by the noise if the signal is scattered on many dimensions than if it is concentrated on few dimensions. This remark highlights the fact that the SNR is not necessarily a good measure of the level of noise in a data set.
In addition, the results of the SURE method are quite poor when the SNR is high.
This can be explained by the fact that the SURE method takes into account too many dimensions (since all the singular values which are higher than the threshold $\lambda$ are kept) in the estimation of $\hat{\bfX}^{\mbox{SURE}}$. For example, with $n = 100$, $p = 20$, $S = 2$, $SNR = 4$ and $(d_1/d_2) = 4$ (first row), the SURE method considers between 9 and 13 dimensions to estimate $\hat{\bfX}^{\mbox{SURE}}$.

Finally, the behaviour regarding the ratio $(n/p)$ is worth noting of. The MSEs are in the same order of magnitude for $(n/p) = 0.2$ and $(n/p) = 5$ and are much smaller for $(n/p) = 1$ for all the methods. 
The issue of dimensionality does not occur only when the number of variables is much larger than the number individuals. 
Rather, difficulties arise when one mode ($n$ or $p$) is larger than the other one, which can be explained by the bilinear form of the model. 

\subsection{Simulations from \citet{candes2012}}

Regularised PCA is also assessed using the simulations from \citet{candes2012}.
Simulated matrices of size $200\times 500$ were drawn with 4 SNR (0.5, 1, 2 and 4) and 2 numbers of underlying dimensions (10, 100). 

Results for the SURE method (Table \ref{candes}) are in agreement with the results obtained by \citet{candes2012}.
As in the first simulation study (section~\ref{sec:simulationStudy}), rPCA outperforms both PCA and the SURE method in almost all cases.
However, the SURE method provides better results than rPCA when the number of underlying dimensions $S$ is high ($S = 100$) and the SNR is small (SNR = 1, 0.5). This is in agreement with the previous comments highlighting the ability of the SURE method to handle noisy situations. 
Nevertheless, we note that when the SNR is equal to 0.5, rPCA is performed with the ``true'' number of underlying dimensions (100). However, if we 
estimate the number of underlying dimensions on these data with one of the available methods in the literature \citep{jolliffe_principal_2002}, all the methods select 0 dimensions.
Indeed, the data are so noisy that the signal is nearly lost. Results obtained with rPCA, using 0 dimensions results in estimating all the values of $\hat\bfX^{\mbox{rPCA}}$ by 0 which corresponds to an MSE equal to 1. 
In this case, considering 0 dimensions in rPCA leads to a lower MSE than taking into account 100 dimensions (MSE = 1.48), but it is still higher than the MSE of the SURE method (0.85).

\begin{table*}[!ht]
\caption{
Mean Squared Error (and its standard deviation) between $\hat{\bfX}$ and $\tilde{\bfX}$ for PCA, regularised PCA (rPCA) and SURE method over 100 simulations.
Results are given for $n = 200$ individuals, $p = 500$ variables, different numbers of underlying dimensions ($S$) and signal-to-noise ratios (SNR).
}
\begin{center}
\begin{tabular}{cc|c|c|c}
$S$	&	SNR	& $MSE(\hat\bfX^{\mbox{PCA}}, \tilde\bfX)$ & $MSE(\hat\bfX^{\mbox{rPCA}}, \tilde\bfX)$ & $MSE(\hat\bfX^{\mbox{SURE}}, \tilde\bfX)$ \\
\hline																						
10	&	4	&	4.31E-03 (7.96E-07) &	\textbf{	4.29E-03	} (7.91E-07) &		8.74E-03	 (1.15E-06)\\
10	&	2	&	1.74E-02 (2.84E-06) &	\textbf{	1.71E-02	} (2.81E-06) &		3.29E-02	 (4.68E-06)\\
10	&	1	&	7.16E-02 (1.25E-05) &	\textbf{	6.75E-02	} (1.15E-05) &		1.16E-01	 (1.59E-05)\\
10	&	0.5	&	3.19E-01 (5.44E-05) &	\textbf{	2.57E-01	} (4.85E-05) &		3.53E-01	 (5.42E-05)\\
100	&	4	&	3.79E-02 (2.02E-06) &	\textbf{	3.69E-02	} (1.93E-06) &		4.50E-02	 (2.12E-06)\\
100	&	2	&	1.58E-01 (8.99E-06) &	\textbf{	1.41E-01	} (7.98E-06) &		1.56E-01	 (8.15E-06)\\
100	&	1	&	7.29E-01 (4.84E-05) &		4.91E-01	 (2.96E-05) &	\textbf{	4.48E-01	} (2.26E-05)\\
100	&	0.5	&	3.16E+00 (1.65E-04) &		1.48E+00	 (1.12E-04) &	\textbf{	8.52E-01	} (3.07E-05) \\																						
\end{tabular}
\end{center}
\label{candes}
\end{table*}

The R \citep{cran} code to perform all the simulations is available on request.


\subsection{Recovery of the graphical outputs}

\begin{figure*}[!ht]
	\center
	\includegraphics[scale = 0.41]{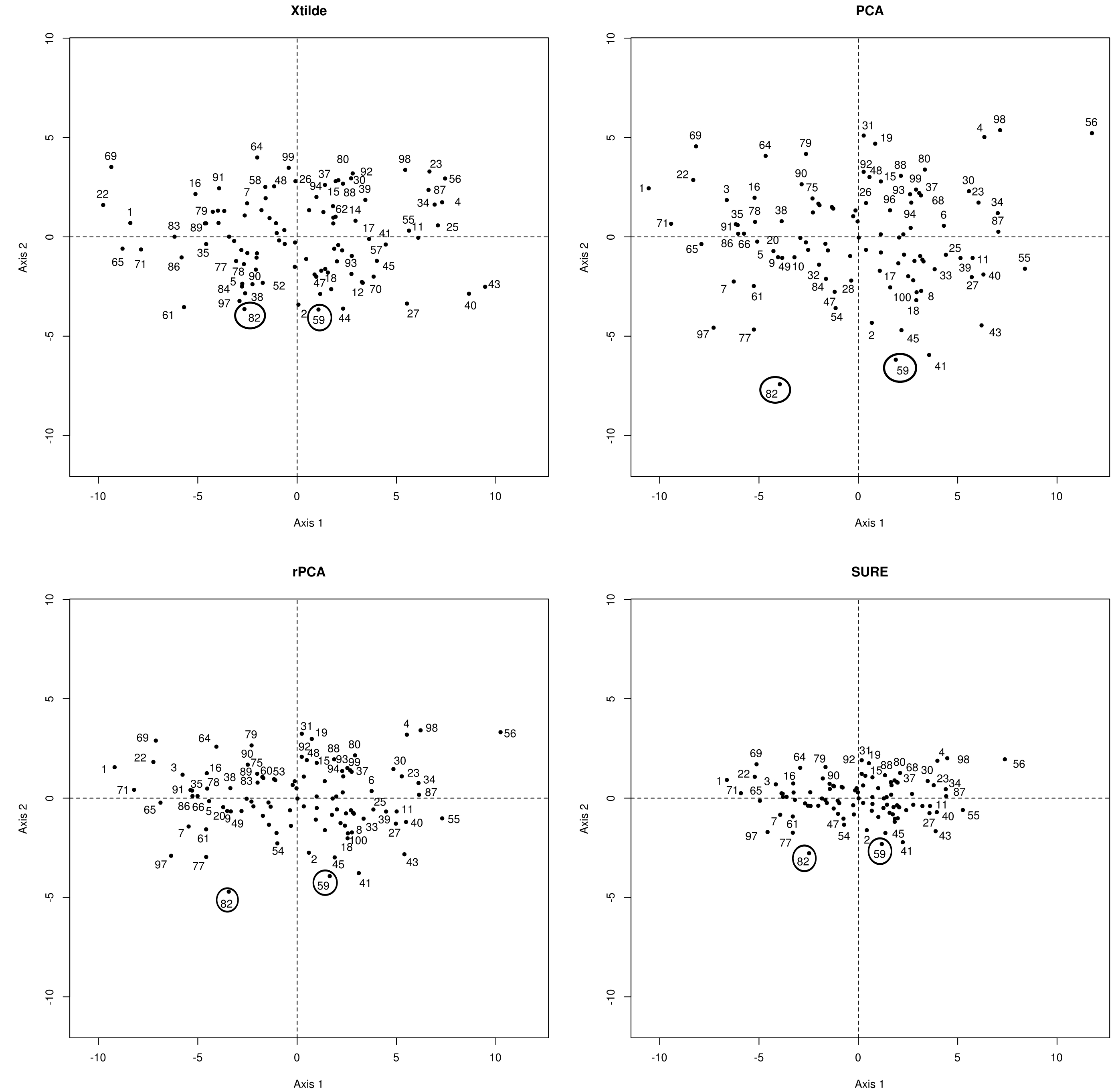}
	\caption{Individual representations of $\tilde{\bfX}$ (top left), of the PCA of $\bfX$ (top right), of the rPCA of $\bfX$ (bottom left) and of the SURE method applied to $\bfX$ (bottom right) for a data set with $n=100$, $p=20$, $S=2$, $(d_1/d_2)=4$ and SNR$=0.8$.}
	\label{ind}
\end{figure*}

Because rPCA better recovers the signal, it produces graphical outputs (individual and variable representations) closer to the outputs obtained from $\tilde{\bfX}$.
We illustrate this point on a simple data set with 100 individuals, 20 variables, 2 underlying dimensions, $(d_1/d_2)=4$ and a SNR equal to 0.8 (row 5 of Table \ref{MSE}). Figure~\ref{ind} provides the true individuals representation obtained from $\tilde \bfX$ (top left) as well as the representations obtained by PCA (top right), rPCA (bottom left) and the SURE method (bottom right). 
The cloud associated with PCA has a higher variability than the cloud associated with rPCA which is tightened around the origin. 
The effect of regularisation is stronger on the second axis than on the first one, which is expected because of the regularisation term.
For instance, the individuals 82 and 59, which have small coordinates on the second axis in PCA are brought closer to the origin in the representation obtained by rPCA which is more in agreement with the true configuration. 
The cloud associated with the SURE method is tightened around the origin on the first axis and even more so on the second one, which is also expected because of the regularisation term. However the global variance of the SURE representation, which is reflected by the variability, is clearly lower than the variance of the true signal.
Therefore, the global shape of the cloud of rPCA is the closest to the true one and thus rPCA successfully recovers the distances between individuals.

\begin{figure*}[!ht]
	\center
	\includegraphics[scale = 0.43]{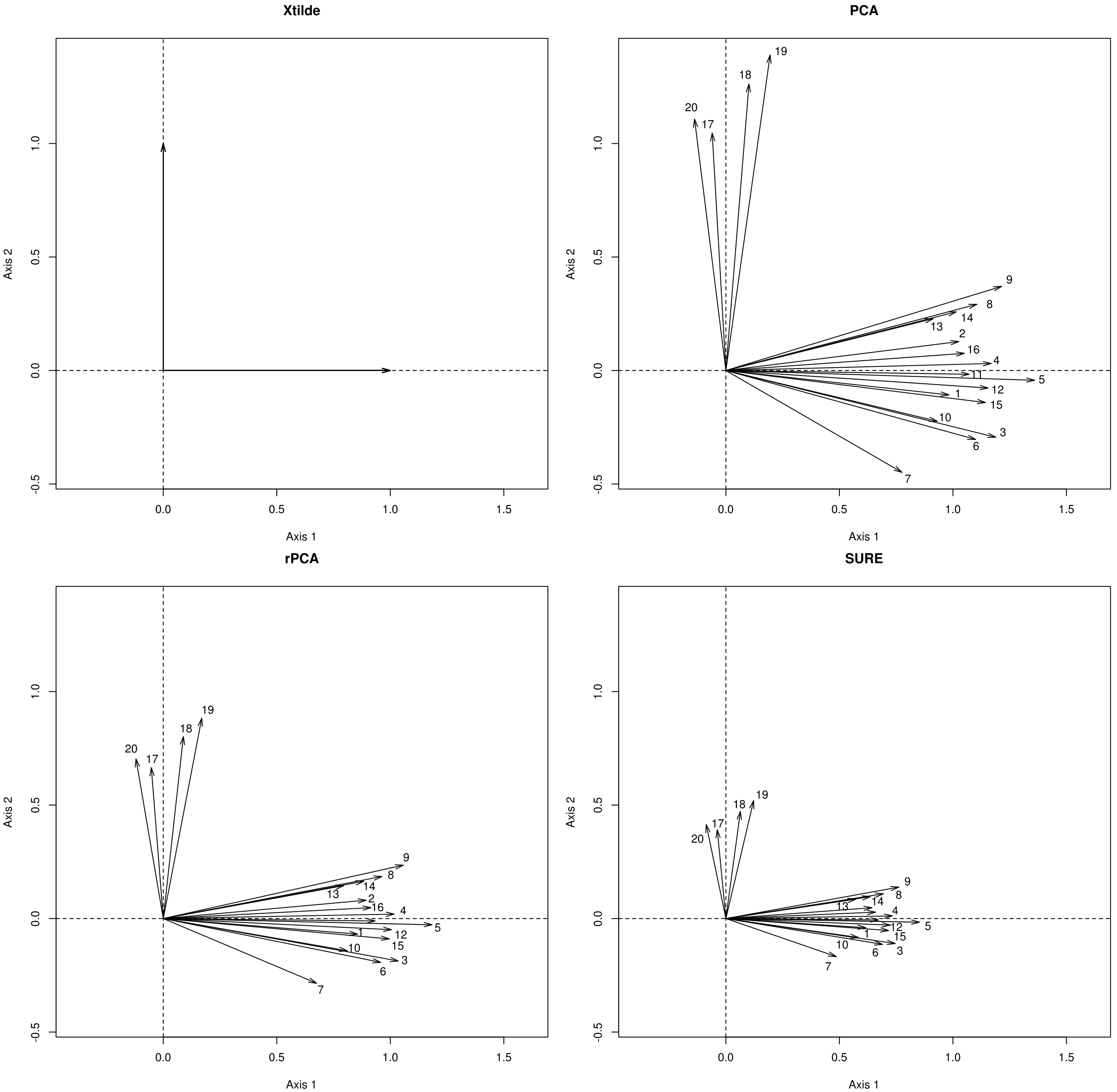}
	\caption{Variable representations of the PCA of $\tilde{\bfX}$ (top left), the PCA of $\bfX$ (top right), the rPCA of $\bfX$ (bottom left) and the SURE method applied to $\bfX$ (bottom right) for an example of data set with $n=100$, $p=20$, $S=2$, $(d_1/d_2)=4$ and SNR$=0.8$.}
	\label{var}
\end{figure*}

Figure~\ref{var} provides the corresponding representations for the variables. The link between the variables which have high coordinates on the first and the second axis of the PCA of $\bfX$ is reinforced in rPCA. This is consistent with the representation of $\tilde \bfX$. For instance, variables 9 and 7 which are correlated to 1 in $\tilde \bfX$ are not very linked in the PCA representation (correlation equal to 0.68) whereas their correlation equals 0.81 in the rPCA representation and 0.82 in the SURE representation.
On the contrary, variables 20 and 7, orthogonal in $\tilde{\bfX}$, have rather high coordinates, in absolute value, on the second axis in the PCA representation (correlation equal to -0.60). Their link is slightly weakened in the rPCA representation (correlation equal to -0.53) and in the SURE representation (correlation equal to -0.51).
In addition, all the variables are generated with a variance equal to 1. 
The variances are over-estimated in the PCA representation and under-estimated in the SURE representation, particularly for the variables which are highly linked to the second axis. 
The best compromise for the variances is provided by rPCA. 
Therefore, rPCA successfully recovers the variances and the covariances of the variables.

This example shows that rPCA is a good method to recover the distances between individuals as well as the links between variables. This property of preserving distances is crucial in clustering for instance, as we will show in the applications (section \ref{sec:application}).


	\section{Applications}
	\label{sec:application}

	\subsection{Transcriptome profiling}
	\label{sec:poulets}


Regularised PCA is applied to a real data set \citep{desert2008} which consists of a collection of 12664 gene expressions in 27 chickens submitted to 4 nutritional statuses:  continuously fed (N), fasting for 16 hours (F16), fasting for 16 hours then refed for 5 hours (F16R5), fasting for 16 hours then refed for 16 hours (F16R16).

\begin{figure*}[!ht]
	\center
	\includegraphics[scale=0.37]{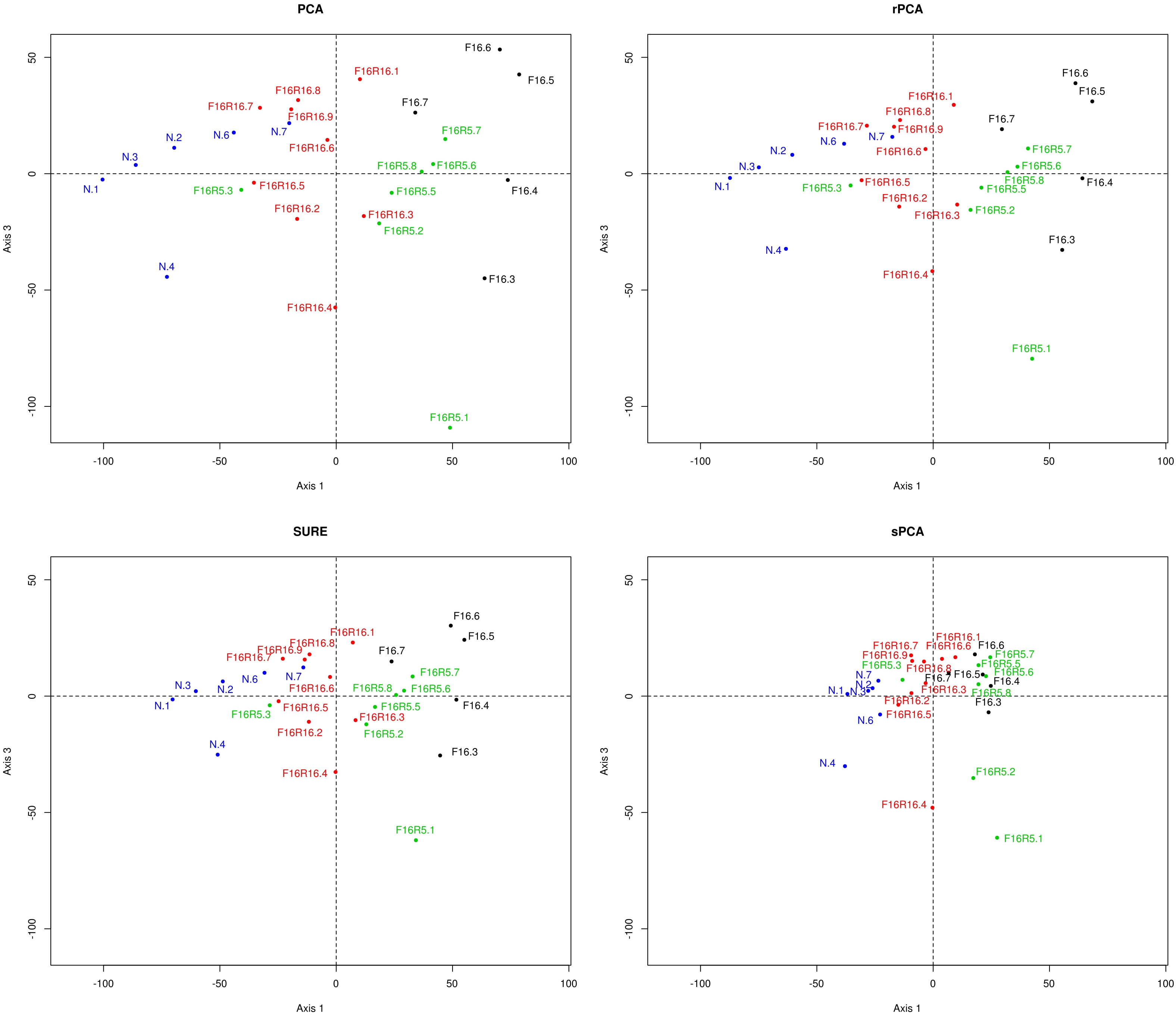}
	\caption{Representation of the individuals on dimensions 1 and 3 of the PCA (top left), the rPCA (top right), the SURE method (bottom left) and sPCA (bottom right) of the transcriptome profiling data. Individuals are coloured according to the nutritional statuses.}
	\label{poule_ind}
\end{figure*}

Since there are 4 nutritional statuses, 3 dimensions are considered. We expect the first three principal components to represent the between-class variability, whereas the following components represent the within-class variability  which is less of interest.
Figure~\ref{poule_ind} shows the individual representations obtained by PCA (top left), rPCA (top right) and the SURE method (bottom left). To better highlight the effect of regularisation,  dimensions 1 and 3 are presented. The first dimension of PCA, rPCA and the SURE method order the nutritional statuses from the continuously fed chickens (on the left) to the fasting chickens (on the right). 
Chickens N.4 and F16R5.1, which have high coordinates in absolute value on the third axis of PCA, are brought closer to the other chickens submitted to the same status in the rPCA representation and in the SURE representation. In addition, chickens N.1 and F16.4, which have high coordinates on the first axis are \linebreak brought closer to the origin in the SURE representation.
Despite these differences, the impact of the regularisation on the graphical outputs appears to be small.

The representation obtained after a sparse PCA (sPCA) method \citep{witten2009} implemented in the R package \texttt{PMA} \citep{PMA} is also provided (bottom right).  Indeed, it is very common to use sparse methods on this kind of data \citep{zou2006}. The basic assumptions for the development of sPCA is that PCA provides principal components that are linear combinations of the original variables which may lead to difficulties during the interpretation especially when the number of variables is very large. 
Loadings obtained via sPCA are indeed sparse, meaning they contain many 0 elements and therefore select only a few variables.
The representation stemming from sPCA is quite different from the other representations; in particular the clusters of F16R5 and of F16 chickens are less clearly differentiated.

\begin{figure*}[!ht]
		\center
	\subfloat[PCA]{\label{heatmap_PCA}\includegraphics[scale=0.27]{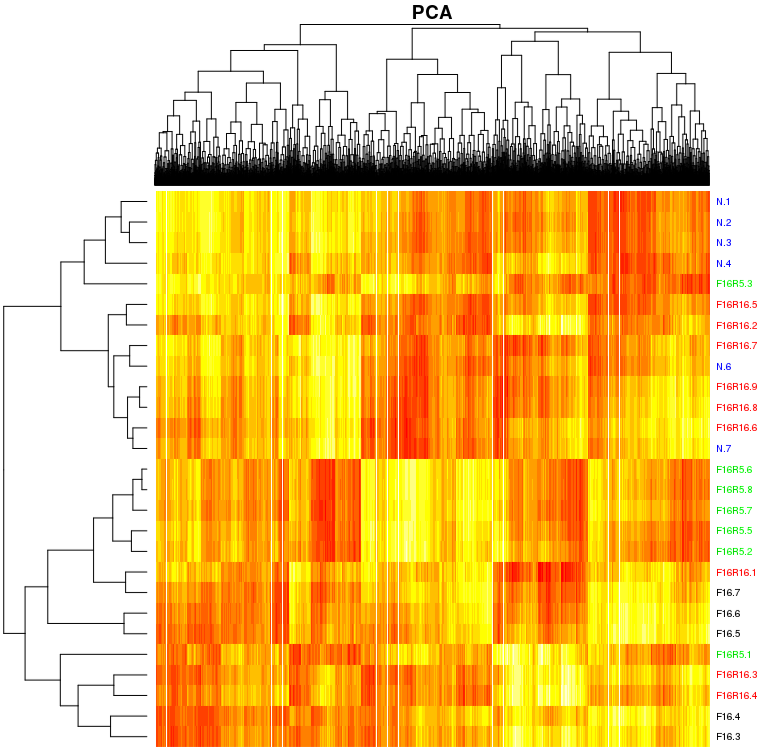}}
	\subfloat[rPCA]{\label{heatmap_rPCA}\includegraphics[scale=0.27]{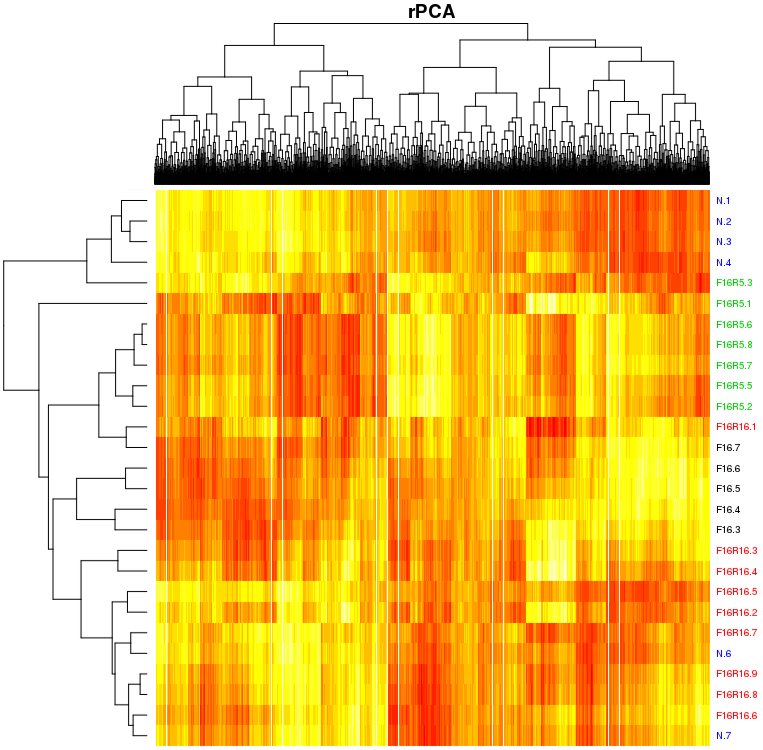}}
	\qquad
	\subfloat[SURE]{\label{heatmap_SURE}\includegraphics[scale=0.27]{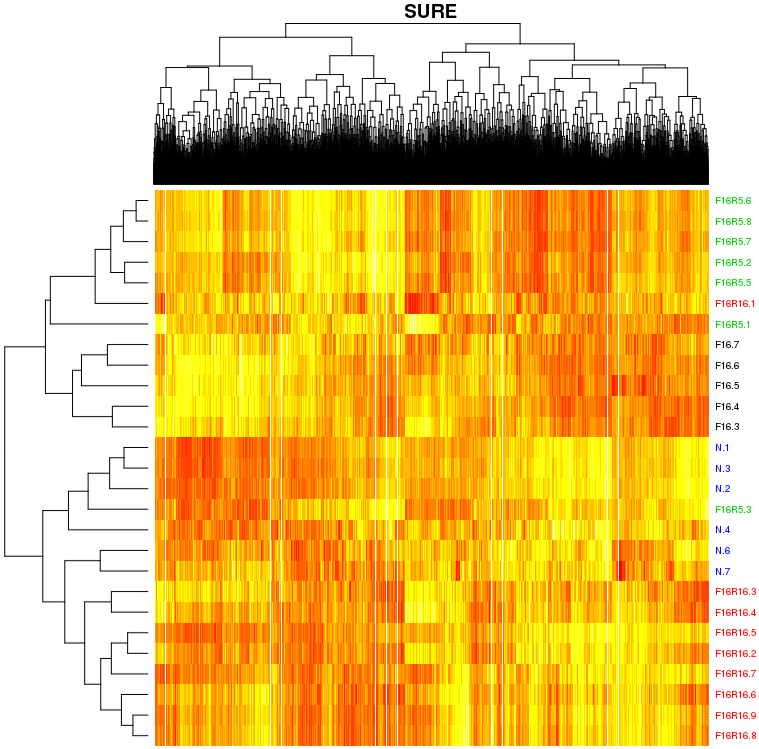}}
	\subfloat[sPCA]{\label{heatmap_sPCA}\includegraphics[scale=0.27]{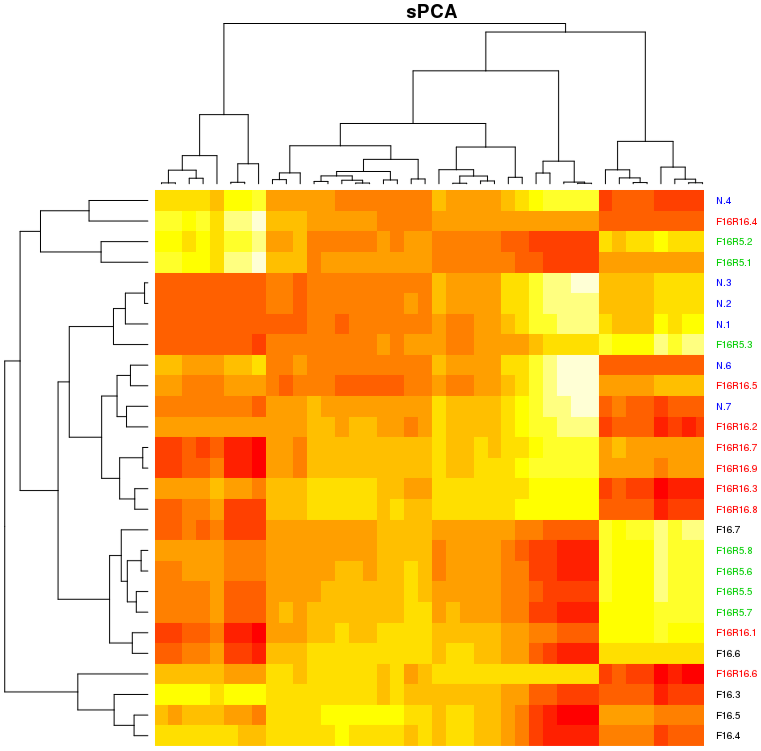}}
	\caption{Heatmaps associated with the analysis of the transcriptomic data set. The data sets used to perform the heatmaps are the fitted matrices stemming from PCA (a),  rPCA (b), the SURE method (c) and sPCA (d).}
	\label{heatmap}
\end{figure*}

It is customary to complement principal components methods with double clustering in order to simultaneously cluster the chickens and the genes and to represent the results using heatmaps \linebreak \citep{eisen1998}. The heatmap clustering is applied to the matrices $\hat \bfX$ obtained by the different methods (Figure~\ref{heatmap}).
Because rPCA modifies the distances between chickens as well as the covariances between genes, the rPCA heatmap will differ from the PCA heatmap.
The rPCA heatmap (Figure \ref{heatmap_rPCA}) is much more appropriate than the PCA heatmap (Figure \ref{heatmap_PCA}).
Indeed, the chickens undergoing 16 hours of fasting are separated into two sub-clusters in the PCA heatmap separated by the chickens F16R5.1, F16R16.3 and  F16R16.4, whereas they are well-clustered in the rPCA heatmap. 
Similarly chickens F16R5 are agglomerated in the PCA heatmap except for chickens F16R5.1 and  F16R5.3,  whereas they are well-clustered in the rPCA heatmap. Finally, the F16R16 chickens are more scattered in both representations. However in rPCA, this can be interpreted as some of the chickens, having fully recovered from the fasting period, are mixed with continuously fed chickens, and some having not fully recovered are mixed with \linebreak F16R5 chickens: the large majority of F16R16 chickens are agglomerated and mixed with N.6 and N.7, and chicken F16R16.1 is mixed with F16R5 chickens. It is not the case for PCA, where the F16R16 chickens are mixed with chickens submitted to all the other nutritional statuses.
The conclusions concerning the SURE heatmap (Figure \ref{heatmap_SURE}) are similar to the conclusions drawn from rPCA. 
The 4 clusters corresponding to the 4 nutritional statuses are well-defined.
However, chicken F16R5.3 is clustered with the N chickens. In addition, the global contrasts are weaker in the SURE heatmap than in the rPCA heatmap.
The heatmap stemming from sPCA (Figure \ref{heatmap_sPCA}) seems to be easier to interpret since there are more contrasts. This is due to the drastic selection of the genes (43 genes were selected among the 12664 genes of the data set). However none of the chicken clusters is clearly defined.

We will not dwell on the interpretation of the gene expressions in the heatmap; however, if the chicken clustering is coherent, the gene clustering is expected to be more coherent as well.

In this example, the impact of regularisation on the graphical representations is not obvious, but the effect of regularisation is crucial to the results of the clustering.
This can be explained by the ability of rPCA to denoise data. 
Such a denoising property can also be useful when dealing with images as illustrated in the next section.


	\subsection{Image denoising}
	\label{sec:image}

We consider the PINCAT numerical Phantom data from \citet{sharif2007} analysed in \citet{candes2012} providing a signal with complex values. 
The PINCAT data simulate a first-pass myocardial perfusion real-time magnetic resonance imaging series, comprising 50 images, one for each time.
To compare the performances of PCA, rPCA and the SURE method, 100 data sets are generated by adding a complex iid Gaussian noise, with a standard deviation equal to 30, to the PINCAT image data. The original image data are then considered as the true (noise-free) images.
PCA and rPCA are performed assuming 20 underlying dimensions. 
This number was chosen empirically and we verified that using slightly more or fewer dimensions does not greatly impact the results.
The SURE method is performed by taking into account the true noise standard deviation which is equal to 30.
The methods are then evaluated by computing the MSE over the 100 simulations. The MSE are equal to 814.26, 598.17 and 727.33 respectively for PCA, rPCA and the SURE method. Consequently, rPCA outperforms both PCA and the SURE method in terms of MSE.

\begin{figure*}[!ht]
	\center
	\includegraphics[scale = 0.6]{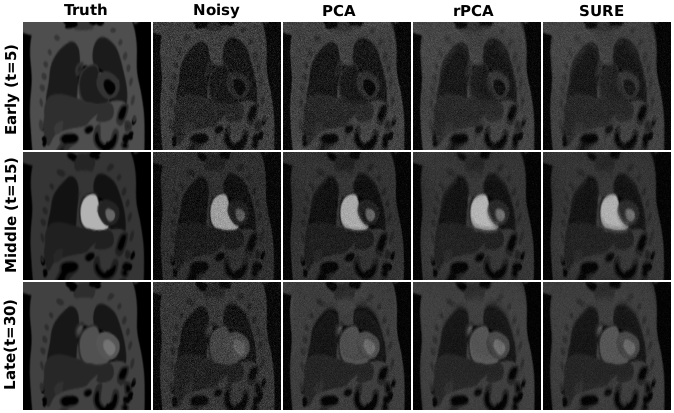}
	\caption{Representation of frames from the PINCAT data at 3 times (early, middle and late) of the true images, the noisy images and the image estimations resulting from PCA, rPCA and SURE.}
	\label{image}
\end{figure*}

In addition, Figure \ref{image} presents a comparison on one simulation of PCA, rPCA and the SURE method. 
Similarly to \citet{candes2012}, we present 3 frames from the PINCAT data (early, middle and late times) for the true image data, the noisy image data, and the image data resulting from denoising by PCA, rPCA and SURE.
All three methods are clearly efficient to reduce the noise; however, the SURE method and rPCA provide images with more contrast than the images provided by PCA. Since rPCA has lower MSE it provides images with a higher degree of noise reduction.

\begin{figure*}[!ht]
	\center
	\includegraphics[scale = 0.5]{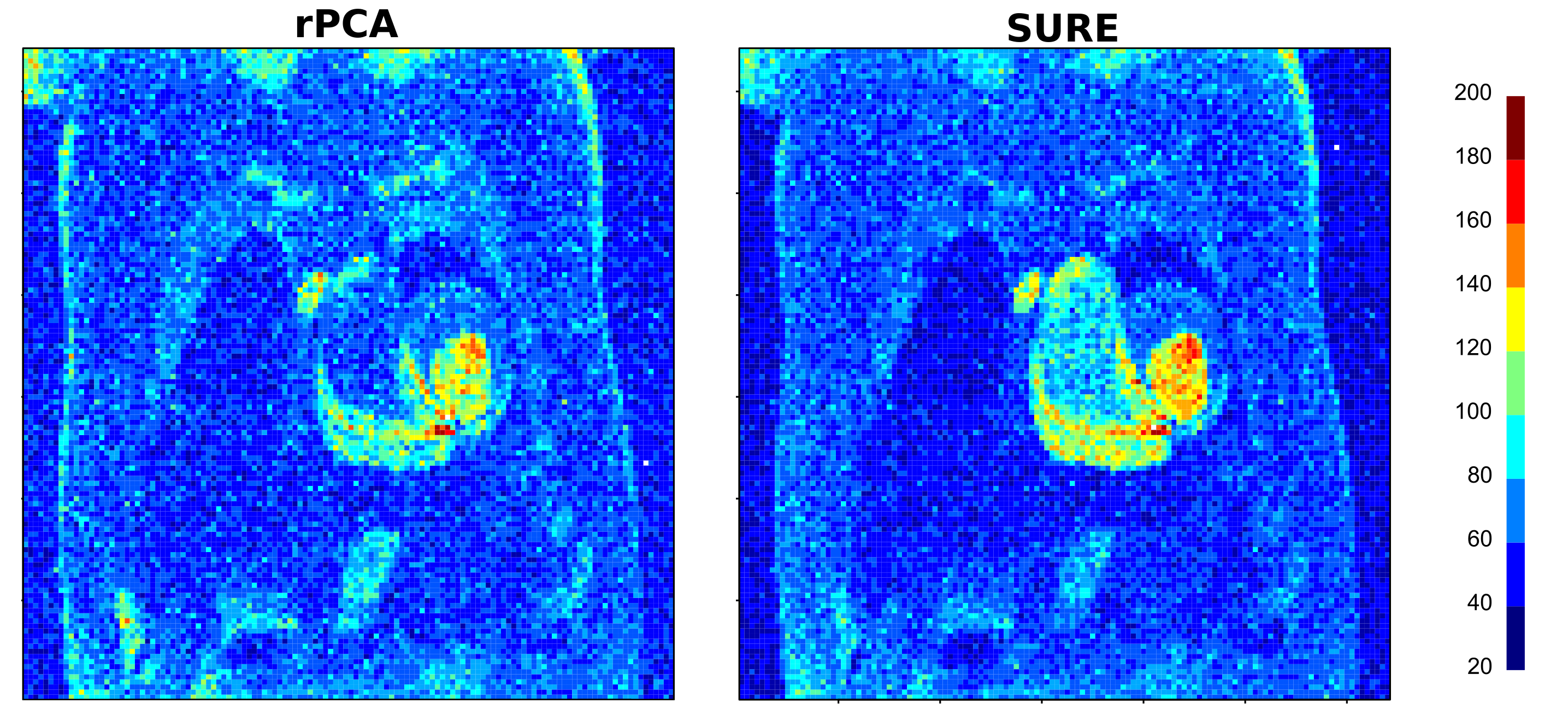}
	\caption{Worst-case absolute error through time of the image estimations by rPCA and SURE.}
	\label{erreurs}
\end{figure*}

In addition, we can consider the worst-case absolute error through time (Figure \ref{erreurs}), which is the highest residual error for each pixel at any time. The SURE method has a particularly high residual error in the area near the myocardium which is an area of high motion. 
The residual error is globally lower for rPCA than for SURE, and it is overall lower in the myocardium area.

Therefore, rPCA is a very promising method to denoise image data.


\section*{Conclusion}

When data can be seen as a true signal corrupted by error, PCA does not provide the best recovery of the underlying signal. Shrinking the singular values improves the estimation of the underlying structure especially when data are noisy. Soft thresholding is one of the most popular strategies and consists in linearly \linebreak shrinking the singular values. The regularised PCA suggested in this paper applies a nonlinear transformation of the singular values associated with a hard thresholding rule. The regularised term is analytically derived from the MSE using asymptotic results from nonlinear regression models or using Bayesian considerations.
In the simulations, rPCA outperforms the SURE method in most situations.
We showed in particular that rPCA can be used beneficially prior to clustering (of individuals and/or variables) or in image denoising.
In addition, rPCA allows improvement on the graphical representations in an exploratory framework.
In this framework, it is worth quoting the work of \citet{takane2006} and \citet{hwang2009} who suggested a regularised version of multiple correspondence analysis, \linebreak which also improves the graphical representations. 

Regularised PCA requires a tuning parameter which is the number of underlying dimensions. Many methods \citep{jolliffe_principal_2002} are available in the literature to select this parameter. However, it is still a difficult problem and an active research area. A classical statement is the following: if the selected number of dimensions is smaller than the rank $S$ of the signal, some of the relevant information is lost and, in our situation, this results in overestimating the noise variance. On the contrary, selecting more than $S$ dimensions appears preferable because all the signal is taken into account even if the noise variance is underestimated. However, in case of very noisy data, the signal is overwhelmed by the noise and is nearly lost. In such a case, it is better to select a number of dimensions smaller than $S$. This strategy is a way to regularise more which is acceptable when data are very noisy. In practice, we use a cross-validation strategy \citep{josse2011} which behaves desirably in our simulations (that is, to find the true number of dimensions when the signal-to-noise ratio is large, and to find a smaller number when the signal-to-noise ratio is small).




\bibliographystyle{spbasic}      

\bibliography{MyBiblio}


\end{document}